\documentclass[a4paper,11pt]{article}
\pdfoutput=1
\usepackage{amsmath,bm,amssymb,color,mathtools}

\usepackage{graphicx}
\usepackage{jheppub} 
\usepackage{xspace}
\usepackage{color}
\usepackage{booktabs}

\renewcommand{\tabcolsep}{1.5em}

\DeclareRobustCommand{\ensuremathrm}[1]{\ensuremath{\mathrm{#1}}\xspace}

\DeclareRobustCommand{\PZ}{{\ensuremathrm{Z}}\xspace}
\DeclareRobustCommand{\PW}{{\ensuremathrm{W}}\xspace}
\DeclareRobustCommand{\PH}{{\ensuremathrm{H}}\xspace}
\DeclareRobustCommand{\rs}{\ensuremathrm{s}}
\DeclareRobustCommand{\rT}{\ensuremathrm{T}}

\DeclareRobustCommand{\alphas}{\ensuremath{\alpha_\rs}\xspace}
\DeclareRobustCommand{\mur}{\ensuremath{\mu_\mathrm{R}}\xspace}
\DeclareRobustCommand{\muR}{\mur}
\DeclareRobustCommand{\muf}{\ensuremath{\mu_\mathrm{F}}\xspace}
\DeclareRobustCommand{\muF}{\muf}

\DeclareRobustCommand{\PH}{{\ensuremathrm{H}}\xspace}
\DeclareRobustCommand{\PZ}{{\ensuremathrm{Z}}\xspace}

\DeclareRobustCommand{\GeV}{\ensuremathrm{GeV}\xspace}

\DeclareRobustCommand{\NNLOJET}{\textsc{NNLOjet}\xspace}
\DeclareRobustCommand{\nnlojetPeople}{Xuan Chen, James Currie, Rhorry Gauld, Aude Gehrmann--De Ridder, Marius H\"ofer, Imre Majer, Tom Morgan, Jan Niehues, Joao Pires, Duncan Walker and James Whitehead\xspace}

\title{Second-order QCD effects in Higgs boson production through vector boson fusion}
\author{J.\ Cruz-Martinez$^a$, T.\ Gehrmann$^b$, E.W.N.\ Glover$^{a}$, A.\ Huss$^c$}
\affiliation{$^a$Institute for Particle Physics Phenomenology, Durham University, Durham, DH1 3LE, UK\\
$^b$Physik-Institut, Universit\"at Z\"urich,
Winterthurerstrasse 190,\\CH-8057 Z\"urich, Switzerland\\
$^c$Theoretical Physics Department, CERN, Geneva, Switzerland}
\emailAdd{j.m.cruz-martinez@durham.ac.uk}
\emailAdd{thomas.gehrmann@uzh.ch}
\emailAdd{e.w.n.glover@durham.ac.uk}
\emailAdd{alexander.huss@cern.ch}
\keywords{QCD, Jets, Collider Physics, NLO and NNLO Calculations}
\abstract{We compute the factorising second-order QCD corrections to the 
electroweak production of a Higgs boson through vector boson fusion. Our calculation  is 
fully differential in the kinematics of the Higgs boson and of the final state jets, and uses the antenna 
subtraction method to handle infrared singular configurations in the different parton-level contributions. 
Our results allow us to reassess the impact of the next-to-leading order (NLO) QCD corrections to 
electroweak Higgs-plus-three-jet production and of the next-to-next-to-leading order (NNLO)
QCD corrections to electroweak Higgs-plus-two-jet production. The NNLO corrections are 
found to be limited in magnitude to around $\pm 5\%$ and are  
uniform in several  
of the kinematical variables, displaying a kinematical dependence only in the transverse momenta 
and rapidity separation of the two tagging jets. 
}
\preprint{{CERN-TH-2018-020, IPPP/18/8, ZU-TH 05/18}}

\begin{document}
\maketitle
\allowdisplaybreaks

\section{Introduction}

The discovery of the Higgs boson at the CERN Large Hadron Collider (LHC)~\cite{higgsexp} has initiated an intensive program of precision measurements of the Higgs boson properties, and of its interactions with all other elementary particles.  
A large spectrum of Higgs boson decay modes and production channels are being
investigated at the LHC.
The Higgs boson can be produced at hadron colliders~\cite{djouadi} either 
through its Yukawa coupling to the top quark  (in gluon fusion through a closed top quark 
loop or by associated production with top quarks) or
through its coupling to the electroweak gauge bosons. This electroweak coupling gives rise to two production modes:  associated production with a vector boson, and vector boson fusion (VBF).  

At LHC energies,  the VBF process is the
second-largest inclusive production mode for Higgs bosons, amounting to about 10\% of the dominant 
gluon fusion process. The detailed experimental study of the VBF production mode probes the 
electroweak coupling structure of the Higgs boson, thereby testing the Higgs mechanism of electroweak 
symmetry breaking. These studies do however require 
that VBF events can be discriminated against other Higgs boson production modes, especially against gluon fusion. This can be 
accomplished by exploiting the 
fact that at leading-order (LO) VBF production proceeds with an initial state configuration of two quarks/anti-quarks each radiating 
a weak vector boson, which then fuse to form the observed Higgs boson. The incoming quarks are deflected and lead to energetic jets at large rapidities. The distinctive VBF 
signature is therefore given by 
Higgs-plus-two-jet production, 
with the jets being strongly separated in  rapidity, and forming a di-jet system of high invariant mass. 
These requirements can be formulated in a set of 
VBF cuts~\cite{Barger:1994zq, Rainwater:1998kj} ensuring an event selection 
that enhances VBF events while suppressing the other production modes. 

Perturbative corrections to Higgs boson production via VBF (electroweak Higgs-plus-two-jet production) 
have been derived at 
next-to-leading order (NLO) in QCD~\cite{Figy:2003nv,Berger:2004pca,Figy:2004pt,Arnold:2008rz}
 and in the electroweak theory~\cite{Ciccolini:2007ec}. To optimise the 
VBF event selection cuts, one would also like to have a reliable description of extra jet activity in the VBF 
process. To this end, NLO QCD corrections have also 
been obtained for electroweak Higgs-plus-three-jet production~\cite{Figy:2007kv,Jager:2014vna,Campanario:2013fsa}.
 Next-to-next-to-leading order (NNLO) 
QCD corrections to the inclusive VBF Higgs production cross section were found to be very small~\cite{Bolzoni:2010xr}, 
they are further improved by third-order (N3LO) corrections~\cite{Dreyer:2016oyx}.
However, more sizable NNLO QCD effects were observed for fiducial cross sections and differential 
 distributions in the VBF Higgs-plus-two-jet production process~\cite{Cacciari:2015jma}. The latter calculation used the 
 NLO QCD  Higgs-plus-three-jet production results of Ref.~\cite{Figy:2007kv} 
 as an input, employing a projection to Born-level 
 kinematics to construct the NNLO differential cross section.

 In this paper, we present an independent derivation  of the second-order QCD corrections to the
 electroweak Higgs-plus-two-jet (VBF-$2j$) production process, and use these to make NLO QCD predictions 
 for  VBF Higgs-plus-three-jet production (VBF-$3j$) and NNLO QCD predictions for VBF-$2j$ production. 
 Both predictions are fully differential in the final state kinematics, and allow the computation of 
 fiducial cross sections and differential distributions. In Section~\ref{sec:method}, we describe the 
 calculational method and its implementation in the \NNLOJET framework. Section~\ref{sec:results} 
 contains numerical results for the cross sections and distributions in the VBF-$3j$ and VBF-$2j$ processes at LHC, 
 and Section~\ref{sec:conc} concludes with an outlook. 

\section{Method}
\label{sec:method}

\begin{figure}[t]
  \centering
  \includegraphics[width=0.3\linewidth]{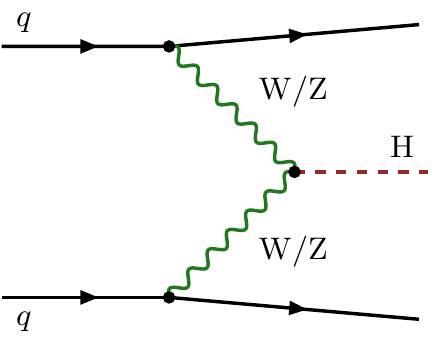}
  \caption{Born-level vector boson fusion process.\label{fig:born}}
  \bigskip
  \includegraphics[width=0.3\linewidth]{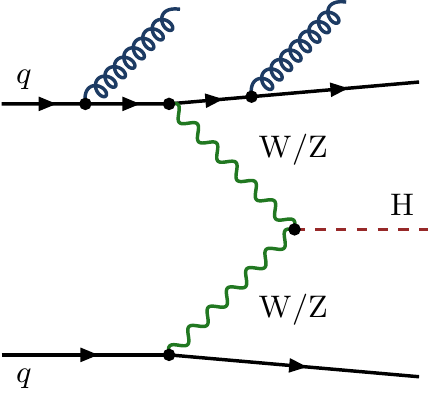}\hfill
  \includegraphics[width=0.3\linewidth]{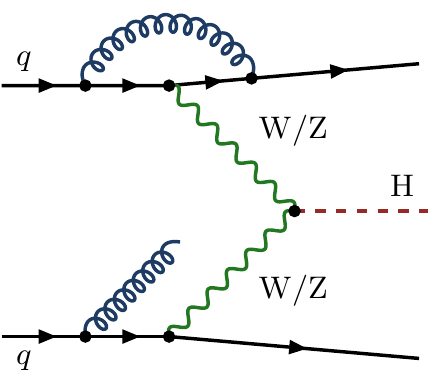}\hfill
  \includegraphics[width=0.3\linewidth]{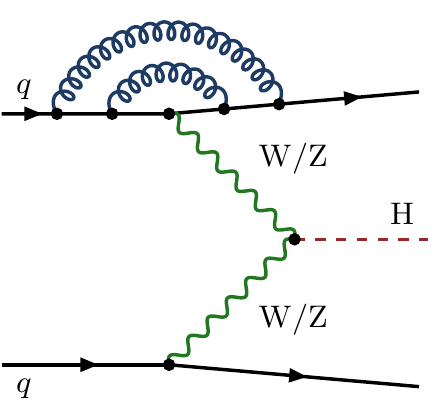}
  \caption{Examples of second order QCD corrections (RR, RV, VV) to the VBF process.\label{fig:nnlofeyn}}
\end{figure}

The Born-level VBF process consists of two independent quark lines, each emitting 
an electroweak gauge boson, linked through a HWW or HZZ vertex, as depicted in Figure~\ref{fig:born}. 
The lack of colour exchange between the two initial state partons means hadronic activity in the 
central region 
is suppressed with respect to other important Higgs production channels, 
where the complicated colour structure means that radiation in the central region is enhanced. Precisely this feature lies at 
the heart of the VBF cuts designed to single 
out VBF over other production modes~\cite{Barger:1994zq, Rainwater:1998kj}. 
Besides enhancing the relative contribution of VBF processes, the VBF cuts also 
strongly suppress interference effects between both quark lines, which are present for 
identical quark flavours. 

When computing higher order QCD corrections, one can exploit this Born-level factorisation of the VBF process into two independent quark lines. Due to color conservation, a single gluon exchange is forbidden between the quark lines, 
such that NLO corrections can be computed by considering corrections to the each quark line independently. 
Since each single quark line in the VBF process is identical to the deeply inelastic scattering (DIS) process of a quark on a vector boson current, this factorisation into two independent processes is also called the ``structure function approach''~\cite{Han:1992hr}. 
Beyond NLO, one can define the structure function approach by forbidding colour exchange between the quark lines. This results in a gauge-invariant subset of diagrams.
Several studies have been performed, showing that the contributions 
that are neglected in the structure function approach are very small in the relevant phase-space regions 
defined by VBF cuts, even if they are sizeable when no cuts are 
used~\cite{Ciccolini:2007ec,Campanario:2013fsa,Bolzoni:2011cu}. Interference 
effects between the VBF production channel and other production channels are also negligible~\cite{Andersen:2007mp}.

Second order QCD corrections constitute of contributions from double real radiation (RR), single real 
radiation at one loop (RV) and two-loop virtual (VV), see Figure~\ref{fig:nnlofeyn}.
Working in the structure function approach, the corrections to the basic VBF process can be distributed amongst the quark lines, e.g.\ a real emission off one quark line and 
 a virtual correction to the other line (as in Figure~\ref{fig:nnlofeyn}) contributes to the RV process. 

In our calculation, we implemented the matrix elements for all relevant parton-level subprocess, and used the antenna subtraction technique~\cite{ourant} to construct 
subtraction terms for the infrared real radiation singularities in the RR and RV contributions so that these contributions are finite over the whole of phase space. The implicit singularities in the subtraction terms are then rendered explicit through integration over the unresolved phase space and then combined 
with the VV contribution to render this contribution also finite 
 and amenable to numerical integration in four space-time dimensions. The numerical 
implementation is performed in the \NNLOJET parton-level event generator framework, which provides the 
phase-space generator, event handling and analysis routines as well as all unintegrated 
and integrated antenna functions~\cite{ant} that are used to construct the subtraction terms. 

\section{Results}
\label{sec:results}

For our numerical computations, we use the NNPDF3.0 parton distribution functions~\cite{Ball:2014uwa} with the 
value of $\alphas(M_{\PZ})=0.118$ at NNLO, and $M_{\PH}= 125~\GeV$, which is compatible with the combined 
results of ATLAS and CMS~\cite{Aad:2015zhl}. Furthermore, we use the following electroweak parameters as input:
\begin{align}
    M_{\PW} &= 80.398~\GeV, & \Gamma_{\PW} &= 2.141~\GeV, \nonumber \\
    M_{\PZ} &= 91.188~\GeV, & \Gamma_{\PZ} &= 2.495~\GeV.   
    \label{eq:ew_param}
\end{align}
Jets are reconstructed using the anti-$k_\rT$ algorithm~\cite{Cacciari:2008gp} with a radius parameter $R = 0.4$, 
and are ordered in transverse momentum. 
The renormalisation and factorisation scales are chosen as suggested in~\cite{Cacciari:2015jma}:
\begin{equation}
    \mu_0^{2} (p_{\rT}^{\PH}) = \frac{M_{\PH}}{2}\sqrt{\left(\frac{M_{\PH}}{2}\right)^2 + \left(p_{\rT}^{\PH}\right)^2}.
    \label{eq:scale_choice}
\end{equation}
In all plots, the uncertainty bands denote the scale uncertainty taking 
$\muR = \muF = \{\frac{1}{2},1,2\}\times\mu_0 $ whereas the error bars in ratios correspond to the statistical uncertainty
of the numerical Monte Carlo integration.

\begin{table}[t]
  \centering
  \begin{tabular}{ @{\enskip}l c c@{\enskip} }
    \toprule
         & $\sigma^{\text{reference}}$ (fb) & $\sigma^{\NNLOJET}$ (fb) \\
    \midrule
    LO   & $ 4032^{+57}_{-69} $             & $4032^{+56}_{-69} $ \\
    NLO  & $ 3929^{+24}_{-23} $             & $3927^{+25}_{-24} $ \\
    NNLO & $ 3888^{+16}_{-12} $             & $3884^{+16}_{-12} $ \\
    \bottomrule
  \end{tabular}
  \caption{The fully inclusive VBF cross section. The uncertainty corresponds to a scale variation of $\mu_F = \mu_R = \left\{\frac{1}{2}, 1, 2\right\}\times\mu_0$ where $\mu_0$ is given in Eq.~\eqref{eq:scale_choice}. Reference results are taken from \cite{Cacciari:2015jma}.\label{table:fully_inclusive}}
\end{table} 
As a validation of our calculation, we compare against the fully inclusive cross 
section~\cite{Bolzoni:2010xr,Cacciari:2015jma} finding very good agreement as shown in 
Table~\ref{table:fully_inclusive}. We would like to point out the substantial technical difference between 
our calculation and the approach used for the total inclusive cross section in Refs.~\cite{Bolzoni:2010xr,Cacciari:2015jma}. 
In both these works, an inclusive NNLO coefficient function for the VBF process is constructed from 
the NNLO coefficient functions for deep inelastic scattering~\cite{Zijlstra:1992qd}, which have already combined 
all parton-level subprocesses of different final state multiplicity through the optical theorem. 
The same method was also used for the N3LO corrections of inclusive VBF~\cite{Dreyer:2016oyx}, using the 
deeply inelastic coefficient functions at this order~\cite{Vermaseren:2005qc} as input. 
The differential 
cross sections in~\cite{Cacciari:2015jma} are subsequently obtained by a projection of all higher-multiplicity final states 
to Born-level kinematics. When evaluated for the total inclusive VBF cross section,~\cite{Cacciari:2015jma}
automatically reproduces the 
result of~\cite{Bolzoni:2010xr} by construction. In our implementation, all subprocesses of different final state 
multiplicity are evaluated separately, using the antenna subtraction terms to regulate real radiation singularities, and 
the inclusive cross section is assembled from the sum of several different contributions. Obtaining the 
total cross section~\cite{Bolzoni:2010xr,Cacciari:2015jma} is therefore a highly non-trivial test of the 
implementation of all individual subprocesses, and of the proper functioning of the subtraction procedure. 

Our implementation of the second-order QCD corrections to the Born-level VBF process allows us to 
compute any infrared-safe observable to this order. In particular, these include the LO predictions for 
electroweak Higgs~+~4~jet production, NLO predictions for Higgs~+~3~jet production and 
NNLO predictions for Higgs~+~2~jet production. Numerical predictions for the latter two processes are discussed 
in detail in the following.

\subsection{NLO corrections to Higgs~+~3~jet production in VBF}
\begin{figure}[t]
  \centering
  \includegraphics[width=0.4\linewidth]{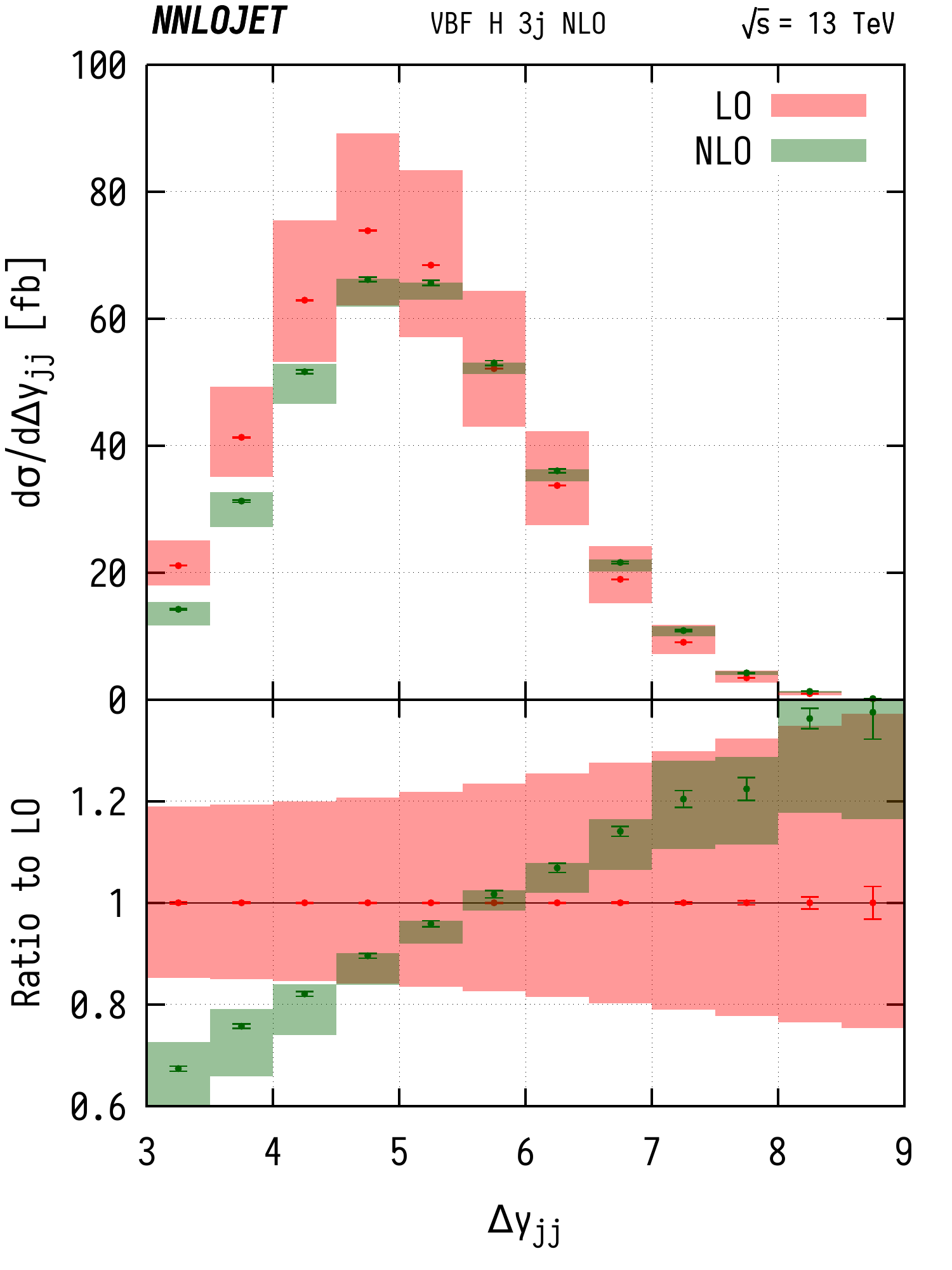}\quad
  \includegraphics[width=0.4\linewidth]{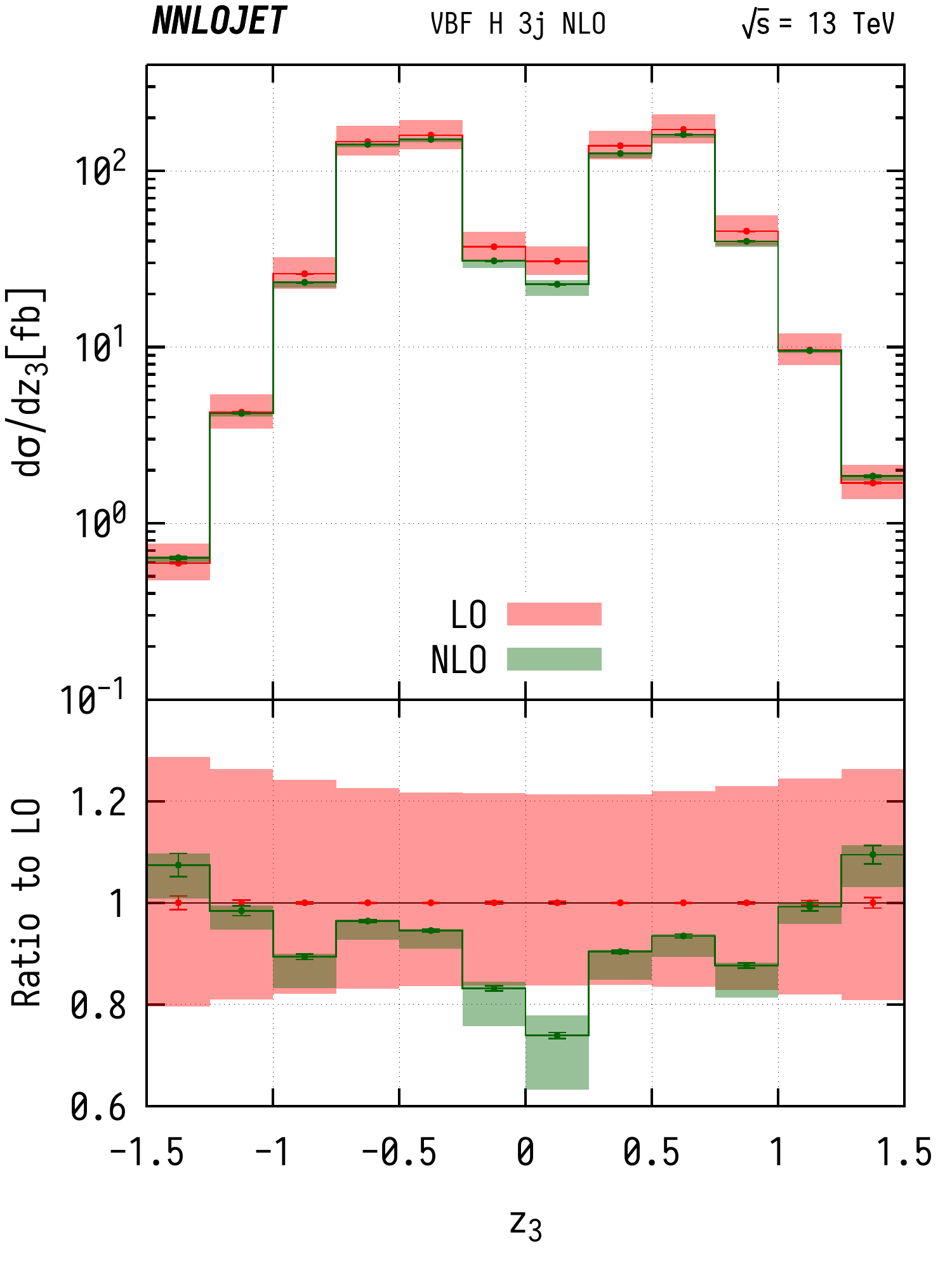}
  \caption{Kinematical distributions  in VBF three-jet process.\label{fig:3jfigure}}
\end{figure}
For the Higgs~+~3~jet production cross section, we require three jets with a transverse
momentum greater than $p_{\rT_j} > 25~\GeV$ and rapidity  $|y_{j}| < 4.5$. Further requirements 
(VBF cuts) are applied to the two leading jets, namely, to their rapidity difference
$\Delta y_{jj}$ and their invariant mass 
$M_{jj}$ to enhance the contribution from the VBF process over other Higgs production mechanisms. 
This leads to the following set of cuts: 
\begin{align}
  p_{\rT_j} &> 25~\GeV, &
  |y_{j}| &< 4.5, \nonumber \\
  M_{jj} &> 600~\GeV, &
  \Delta y_{jj} = | y_{j_1} - y_{j_2} | &> 3, &
  y_{j_1} \cdot y_{j_2} &< 0.
  \label{eq:vbf3j_cuts}
\end{align}
Note that the cut on $ \Delta y_{jj}$ is lower than what would  be required 
to be in a VBF-dominated kinematical region. It has been chosen to allow us to compare our results 
with~\cite{Figy:2007kv} over a larger range in $ \Delta y_{jj}$. 

Figure~\ref{fig:3jfigure} shows the rapidity separation of 
the two leading jets $\Delta y_{jj} = |y_{j_1}-y_{j_2}|$ (left frame) and the normalised 
rapidity distribution of the third jet $z_3 = (y_{j_3}-(y_{j_1}+y_{j_2})/2)/(y_{j_1}-y_{j_2})$ (right frame). In contrast to the 
initial findings of~\cite{Figy:2007kv}, we observe an increase of the NLO corrections for large values 
of $\Delta y_{jj}$. This finding has led to the identification of an error in the virtual matrix elements in~\cite{Figy:2007kv}, and we are in full agreement with the revised results~\cite{figypriv}.

\subsection{NNLO corrections to Higgs~+~2~jet production in VBF}

In the fully inclusive VBF cross section discussed above (Table~\ref{table:fully_inclusive}), we observed an excellent perturbative convergence with very small 
NLO and NNLO corrections and a sizeable reduction of scale uncertainty with each order. 
The inclusive VBF cross section is however not a directly measurable quantity, but only one contribution to
 inclusive Higgs boson production. To single out the VBF contribution, a set of cuts is applied to define 
 VBF-$2j$ production. These are: 
\begin{align}
  p_{\rT_j} &> 25~\GeV, &
  |y_{j}| &< 4.5, \nonumber \\
  M_{jj} &> 600~\GeV, &
  \Delta y_{jj} = | y_{j_1} - y_{j_2} | &> 4.5,
  \label{eq:vbf_cuts}
\end{align}
which are identical to those used in~\cite{Cacciari:2015jma}. A third jet can be present in the event at any rapidity, i.e.\ the cuts define a VBF-$2j$ inclusive cross section. We note that the cut on $\Delta y_{jj}$ is more restrictive than 
in the VBF-$3j$ study in the previous section, and automatically implies that the jets are in opposite hemispheres. 

By application of these cuts, we obtain the fiducial VBF-$2j$ cross sections as listed in Table~\ref{table:fiducial}. 
It is important to note the increase in magnitude of the higher order QCD corrections when 
VBF cuts are applied: we find a negative 
correction factor at both NLO and NNLO which is three times larger in magnitude than
what was found in the inclusive VBF cross section reported in Table~\ref{table:fully_inclusive}. 
\begin{table}[t]
  \centering
  \begin{tabular}{ @{\enskip}l c@{\enskip} }
    \toprule
         & $\sigma^{\NNLOJET}$ (fb) \\
    \midrule
    LO   & $ 957^{+66}_{-59} $  \\
    NLO  & $ 877^{+ 7}_{-17} $  \\
    NNLO & $ 844^{+ 9}_{- 9} $  \\
    \bottomrule
  \end{tabular}
  \caption{Total VBF-$2j$ cross section after cuts are applied.\label{table:fiducial}}
\end{table}

\begin{figure}[t]
  \centering
  \includegraphics[width=0.4\linewidth]{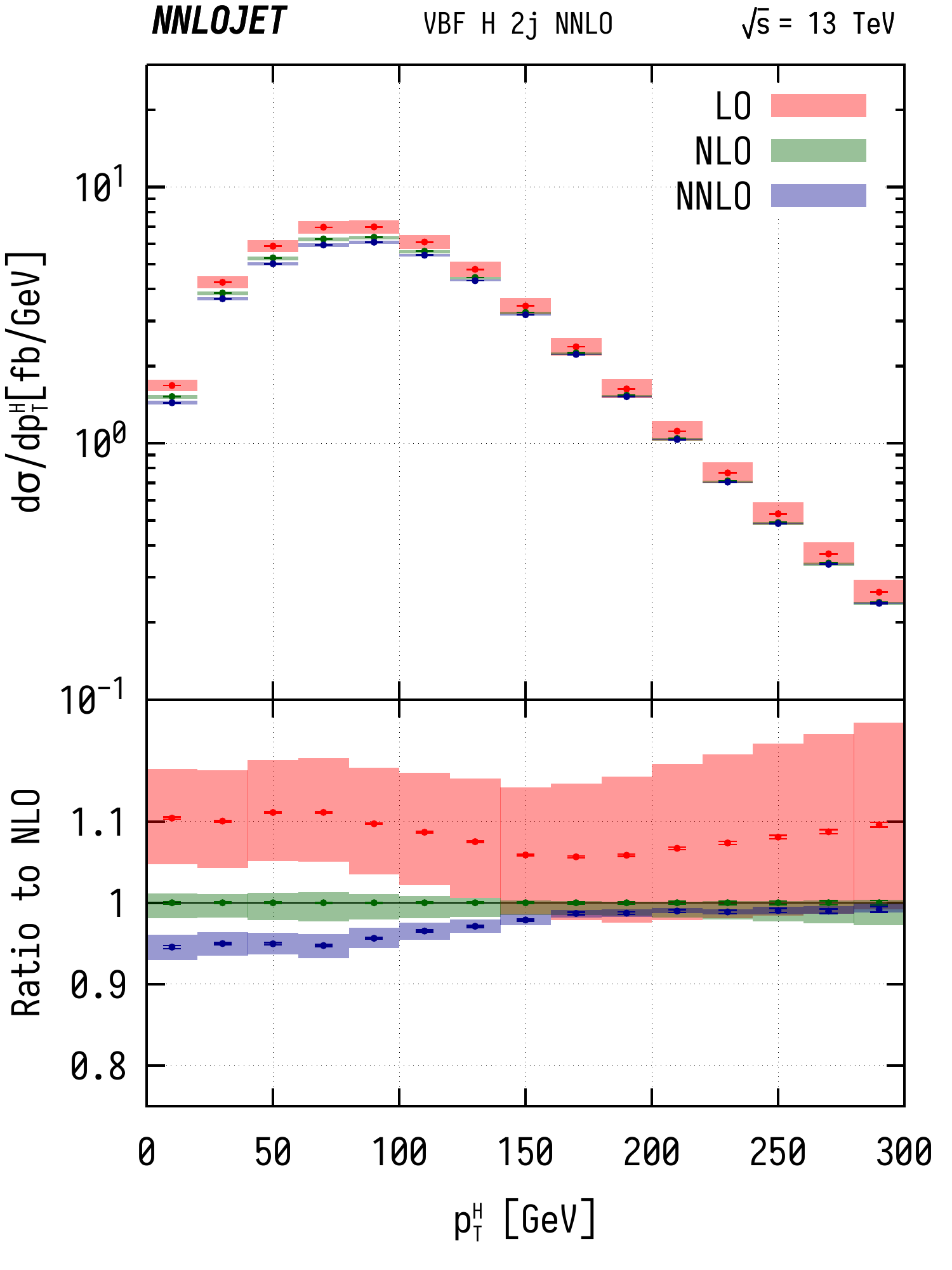}
  \caption{Higgs boson transverse momentum distribution in VBF process.\label{fig:pth}}
  \bigskip
  \includegraphics[width=0.4\linewidth]{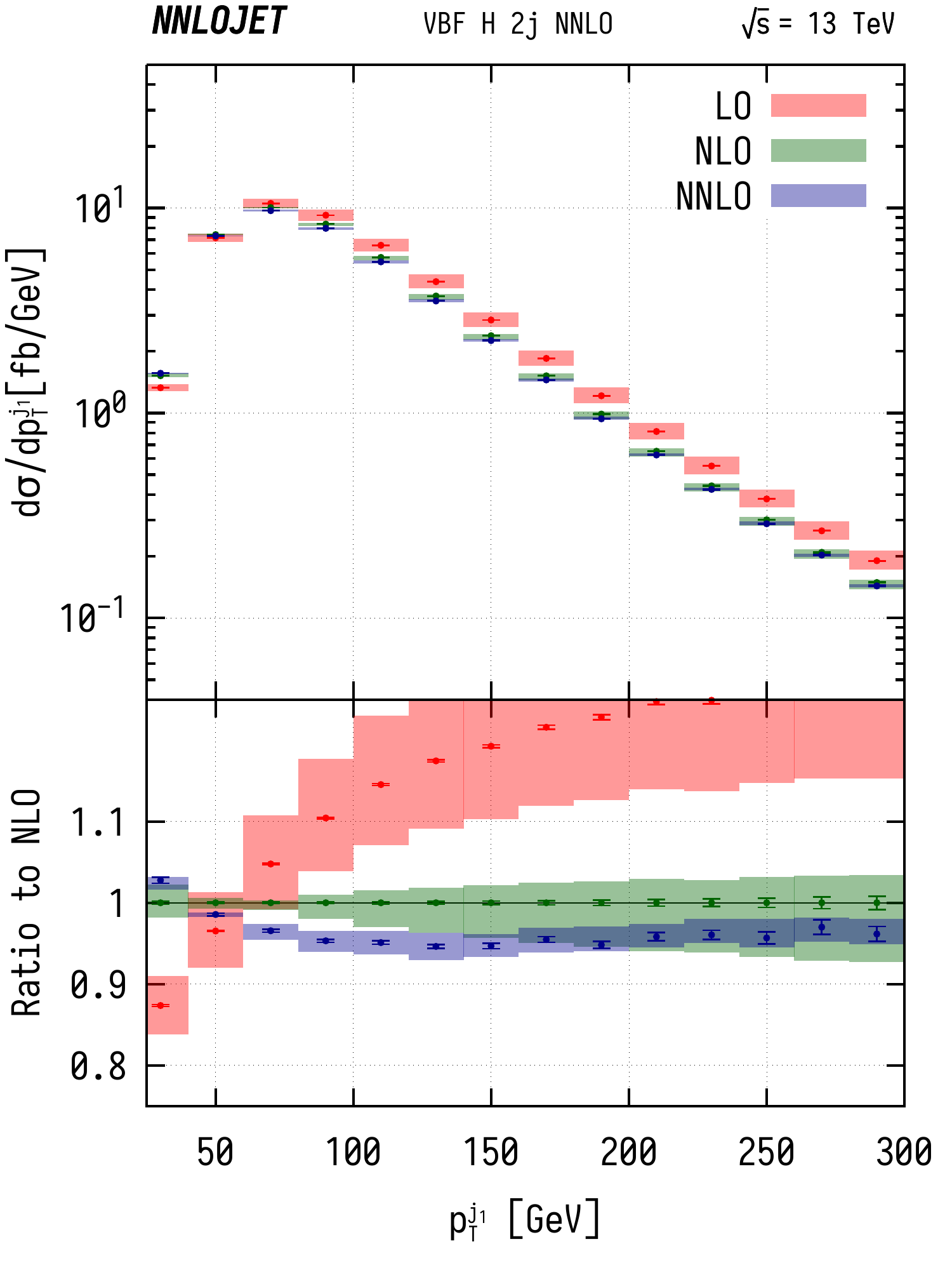}\qquad
  \includegraphics[width=0.4\linewidth]{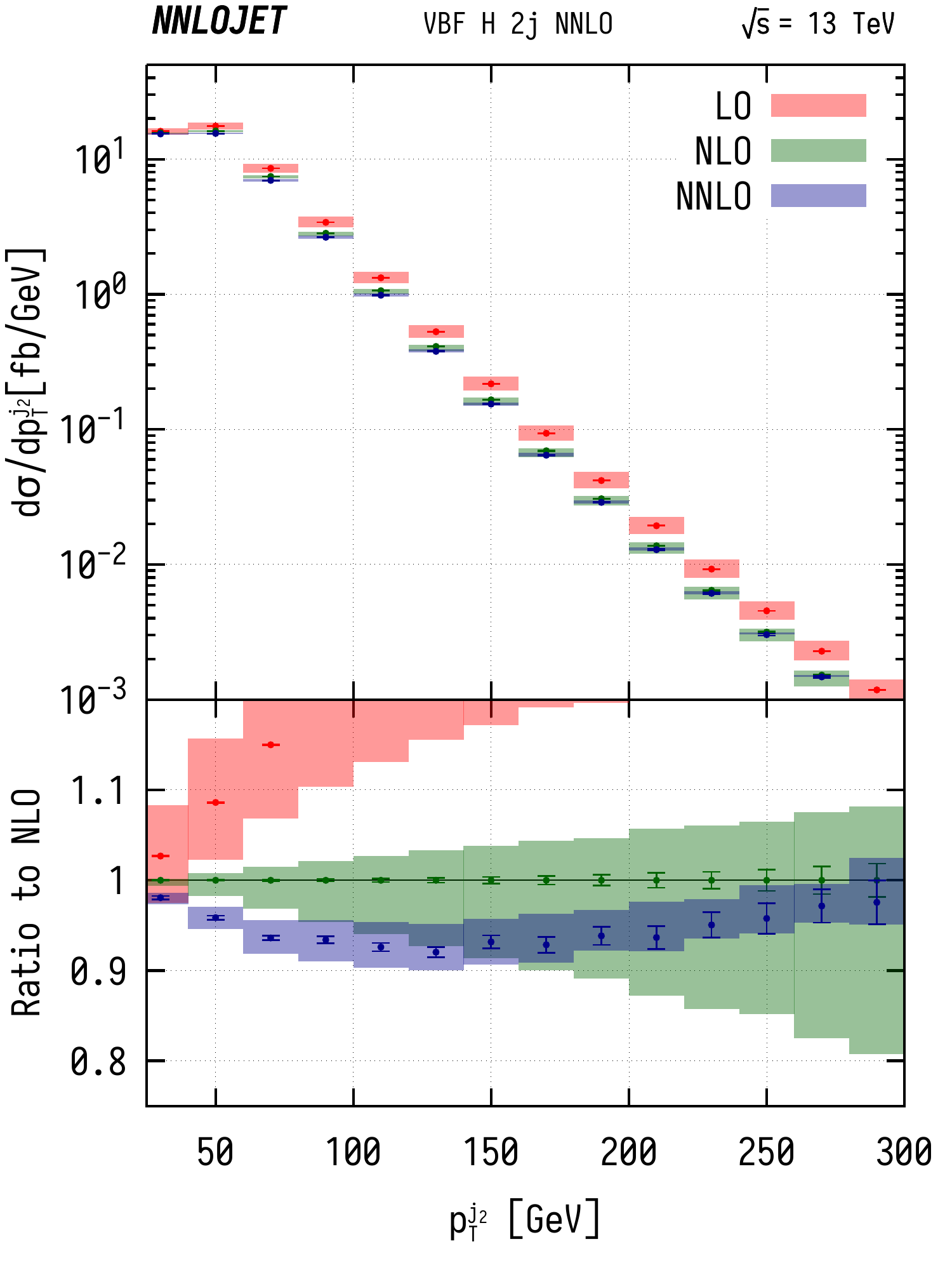}
  \caption{Transverse momentum distribution of leading and subleading jet in VBF process.\label{fig:ptj}}
\end{figure}

The larger impact of the NNLO corrections for the VBF-$2j$ process can also be observed in the differential distributions. 
Figure~\ref{fig:pth} shows the transverse momentum of the Higgs boson. 
The NLO corrections are uniform and negative, amounting to about $-10\%$ throughout the distribution. 
For medium or large transverse momentum, the NNLO correction is quasi-negligible and 
lies within the NLO scale uncertainty band. At lower transverse momentum, where the bulk of the distribution is located, 
 the NNLO corrections become significant at $-5\%$, and lie outside the NLO uncertainty band. 

The transverse momentum distributions of the leading and subleading jet (i.e.\ the two tagging jets for the VBF cuts) are shown in~\ref{fig:ptj}. We observe that the NLO and 
NNLO corrections are both 
less uniform, changing from positive (for the leading jet) or negligible (for the subleading jet) to negative for larger transverse momenta. We also observe that the NLO and NNLO uncertainty bands overlap 
in the range of the observable beyond the very low $p_\rT$ region.
The magnitude of the NNLO corrections is moderate, and never exceeds 5\%, while NLO corrections can be as large as 30\% and lead to a substantial modification of the shape of both jet distributions. 
\begin{figure}[ht]
  \centering
  \includegraphics[width=0.4\linewidth]{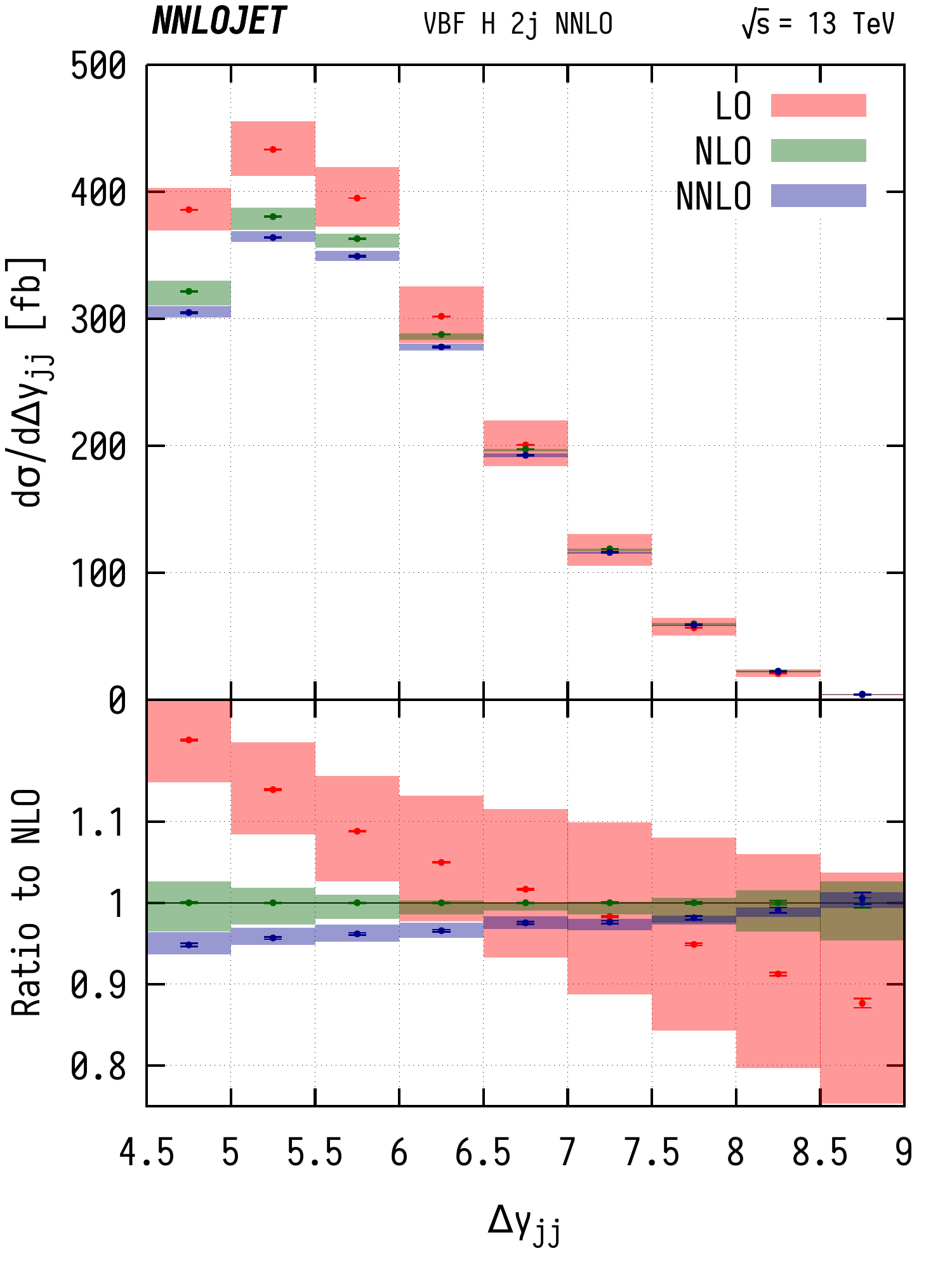}\qquad
  \includegraphics[width=0.4\linewidth]{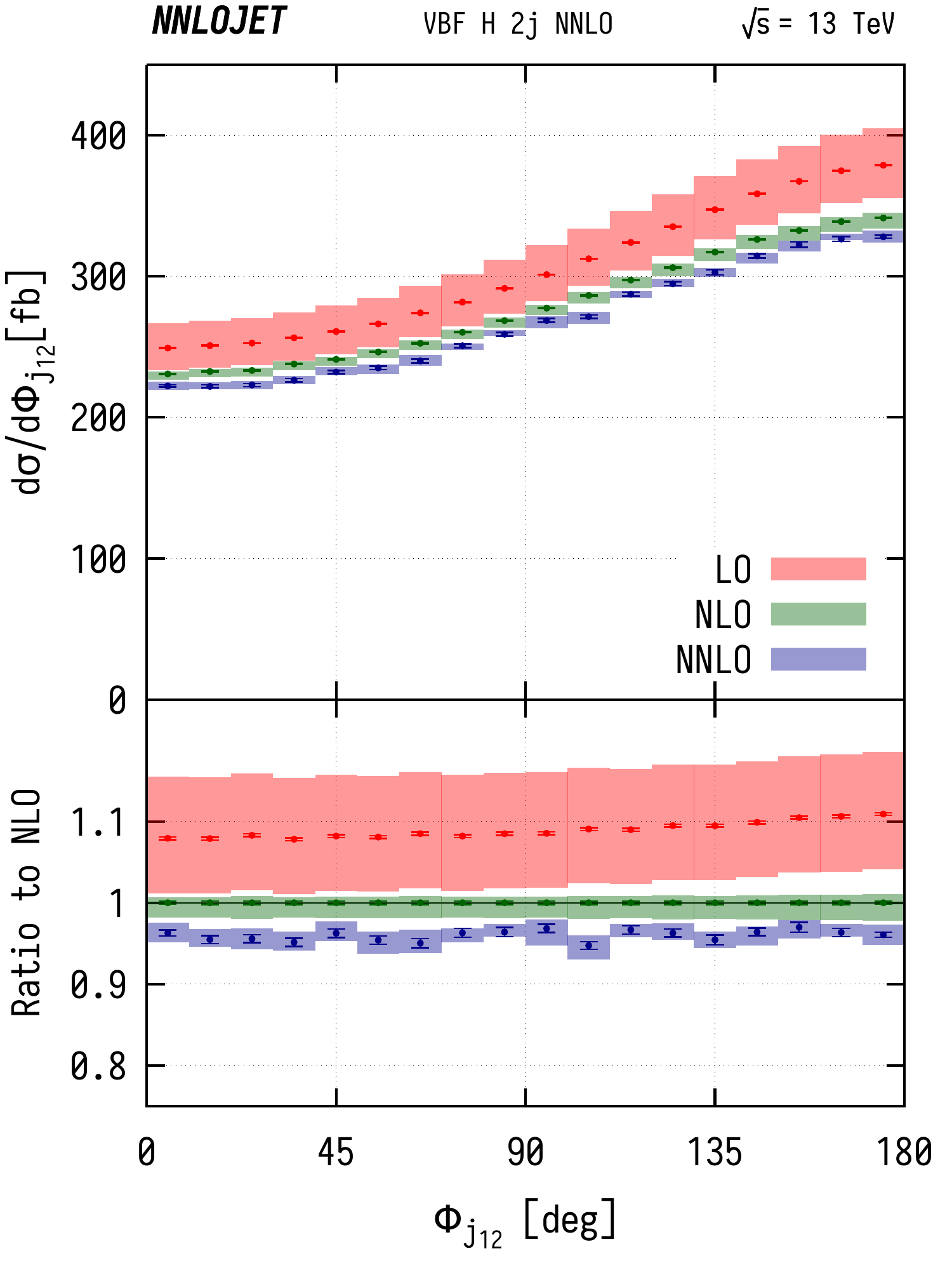}
  \caption{Rapidity separation and angular decorrelation of the two leading jets in the VBF process.\label{fig:deltay}}
  \bigskip
  \includegraphics[width=0.4\linewidth]{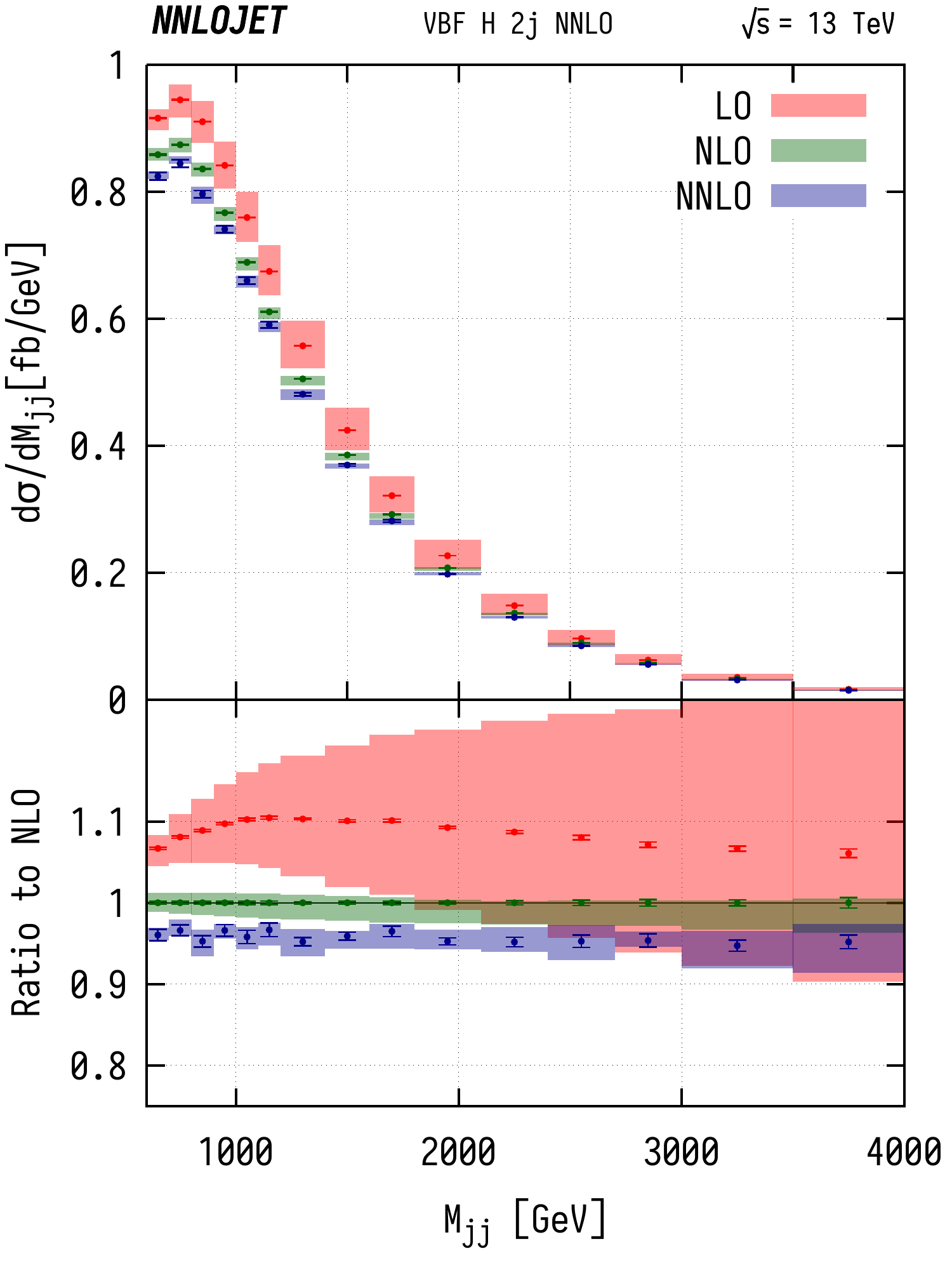}
  \caption{Invariant mass distribution of the two leading jets in VBF process.\label{fig:mj12}}
\end{figure}

The spatial distribution of the two tagging jets is described by their separation in rapidity $\Delta y_{jj}$ and 
their angular decorrelation $\Phi_{j_{12}}$. The VBF-$2j$ distributions in these two variables 
are shown in Figure~\ref{fig:deltay}. We observe that the NLO and NNLO corrections are very uniform in 
$\Phi_{j_{12}}$, while displaying a sizeable dependence on $\Delta y_{jj}$. For low values of this variable (which starts only at $\Delta y_{jj}=4.5$ due to the VBF cuts~(\ref{eq:vbf_cuts})) the corrections are 
 negative and amount to $-25\%$ at NLO and with a further  
$-5\%$ at NNLO. The corrections decrease in magnitude with increasing rapidity separation, and cross zero around $\Delta y_{jj}\sim 7$. 
At even higher separation, the corrections become positive, but remain rather moderate. For both spatial distributions, 
we observe that the NLO and NNLO uncertainty bands barely overlap. Nevertheless, the small magnitude of the NNLO 
corrections indicates a good perturbative convergence. Similar observations also apply to the invariant 
mass distribution of the two tagging jets shown in Figure~\ref{fig:mj12}, where the corrections are also observed to be very uniform. 

The NNLO QCD corrections of VBF-$2j$ production were first computed in~\cite{Cacciari:2015jma}, using 
the Projection-to-Born method applied to the NLO VBF-$3j$ calculation of Ref.~\cite{Figy:2007kv}. 
The recent revision of the latter results~\cite{figypriv} has a significant impact on the VBF-$2j$ distributions in $\Delta y_{jj}$ and $p_{\rT}^{j_2}$. 
Once these corrections are applied~\cite{karlbergpriv} in~\cite{Cacciari:2015jma}, 
we find excellent  agreement with our results for 
the fiducial cross section, Table~\ref{table:fiducial}, and 
all distributions considered in~\cite{Cacciari:2015jma}.

\section{Conclusions}
\label{sec:conc}

We computed the second-order QCD corrections to the electroweak production of a Higgs boson
through the VBF process. Our calculation is  
restricted to corrections that factorise onto either of the two quark lines present in the Born-level process. This approach has 
been proven to be very reliable once VBF cuts are applied. Our results are implemented in the \NNLOJET 
framework, and can be used to compute any infrared-safe observable derived from the VBF process up to 
${\cal O}(\alphas^2)$. Our work provides a  critical validation of earlier results on the NLO QCD corrections 
to VBF-$3j$ production~\cite{Figy:2007kv,figypriv} and the NNLO QCD corrections to VBF-$2j$ 
production~\cite{Cacciari:2015jma,karlbergpriv}. 

The second-order corrections are found to be uniform in most of the kinematical variables, and usually amount to no more
than $5\%$.  We observe a kinematical dependence of the NNLO corrections only in the distributions of the two leading jets (tagging jets) in transverse momentum 
and rapidity separation. Since it is precisely 
through cuts on these variables that the VBF cross section is selected, the NNLO effects may have an important impact on the precise efficiency of the VBF cuts, and consequently on all future precision studies of VBF Higgs boson production.

\section*{Acknowledgements}
The authors thank \nnlojetPeople for useful discussions and their many contributions to the \NNLOJET\ code 
and 
Jonas Lindert for many useful discussions and help in the validation of the relevant matrix elements.
We thank Terrance Figy~\cite{figypriv}, Michael Rauch and Alexander Karlberg~\cite{karlbergpriv} 
for their help in comparing with~\cite{Figy:2007kv} and~\cite{Cacciari:2015jma} which ultimately led to the isolation and rectification of an error in the original code of~\cite{Figy:2007kv}.
We gratefully acknowledge the computing resources provided by the WLCG through the GridPP Collaboration.
This research was supported in part by the National Science Foundation under Grant NSF PHY11-25915,
     by the Swiss National Science Foundation (SNF) under contracts 200020-175595
  and CRSII2-160814, by the Research Executive Agency (REA) of the European Union under the Grant Agreement PITN-GA-2012-316704 (``HiggsTools'') and the ERC Advanced Grant MC@NNLO (340983).

\end{document}